\documentclass[aps,prl,twocolumn,superscriptaddress]{revtex4}

\usepackage{graphicx}
\usepackage{dcolumn}
\usepackage{bm}

\begin{document}
\title{A new chiral electro-optic effect: Sum-frequency generation from optically active liquids in the presence of a dc electric field}
\author{P. Fischer}
\email[Corresponding author: ]{pf43@cornell.edu}
\affiliation{Department of Chemistry and Chemical Biology, Cornell
University, Ithaca, NY 14853, USA} \affiliation{Department of
Applied and Engineering Physics, Cornell University, Ithaca, NY
14853, USA}
\author{K. Beckwitt}
\affiliation{Department of Applied and Engineering Physics,
Cornell University, Ithaca, NY 14853, USA}
\author{A.D. Buckingham}
\affiliation{Department of Chemistry, University of Cambridge, Lensfield Road, Cambridge CB2 1EW, UK}
\author{D.S. Wiersma}
\affiliation{European Laboratory for Non-linear Spectroscopy and
INFM, Via Nello Carrara 1, 50019 Sesto-Fiorentino, Italy}
\author{F.W. Wise}
\affiliation{Department of Applied and Engineering Physics,
Cornell University, Ithaca, NY 14853, USA}

\date{\today}
\begin{abstract}
We report the observation of sum-frequency signals  that depend
linearly on an applied electrostatic field and that change sign
with the handedness of an optically active solution. This recently
predicted chiral electro-optic effect exists in the
electric-dipole approximation. The static electric field gives
rise to an electric-field-induced sum-frequency signal (an achiral
third-order process) that interferes with the chirality-specific
sum-frequency at second-order. The cross-terms linear in the
electrostatic field constitute the effect and may be used to
determine the absolute sign of second- and third-order nonlinear
optical susceptibilities in isotropic media.
\end{abstract}

\maketitle

Most biological molecules are chiral \cite{Kelvin}, i.e. have a
non-superimposable mirror image. In many respects the two mirror
image forms (enantiomers) of a chiral molecule have identical
physical properties. Optical activity is often the only practical
means to distinguish between enantiomers in solution. Conventional
optical activity phenomena, such as optical rotation and circular
dichroism, are based on the interference of induced oscillating
electric- and magnetic (and electric-quadrupole) moments, and
arise from a differential response to left and right circularly
polarized light \cite{Barron}.

Remarkably, a purely electric-dipolar nonlinear-optical process
can also be a probe of chirality \cite{jag1}. Two optical fields
of different frequency can coherently mix in an isotropic medium
to generate light at their sum (or difference) frequency if the
medium is optically active, i.e. a non-racemic solution of chiral
molecules \cite{jag1}. The signal itself is then a probe of
molecular chirality: It is the intrinsic broken symmetry of the
chiral molecules that causes a non-racemic liquid to be
noncentrosymmetric and hence allows for an electric-dipolar
second-order nonlinear optical process in an isotropic medium.

The sum-frequency intensity is in general, however, not sensitive
to the sign of the underlying property tensor, and thus does not
readily distinguish between enantiomers. In this Letter we report
the observation of a recently predicted chiral electro-optic
effect \cite{pfcpl98} that arises when a static electric field is
applied to coherent sum- or difference-frequency generation in an
optically active liquid. The static field does not change the
phase matching conditions, but it gives rise to an
electric-field-induced contribution to the signal. The beat
between chirality-sensitive sum-frequency generation (a
second-order process) and achiral electric-field-induced
sum-frequency generation (a third order process) yields a
contribution to the intensity that is linear in the static
electric field and that changes sign with the enantiomer
\cite{pfcpl98,pfacs01a}. The effect can give the absolute sign of
the isotropic part of the sum-frequency hyperpolarizability (if
the sign of the achiral third-order property is known) and hence
makes it possible to determine the handedness of chiral molecules
in solution via an electric-dipolar optical process.

Sum-frequency generation (SFG) from chiral liquids has recently
been re-examined \cite{pfprl00} and has been observed
experimentally \cite{yrschiral,yrschiral2,pfcpl02}. The molecular
response to two optical fields $E_\beta(\omega_1)$ and
$E_\gamma(\omega_2)$ (and a static field $E_\delta(0)$) at the
sum-frequency $\omega_3=\omega_1+\omega_2$ can be written in terms
of an induced oscillating dipole moment
\begin{eqnarray}
\mu_\alpha(\omega_3)&=&\frac{1}{2} \, \beta_{\alpha \beta
\gamma}(\omega_3=\omega_1+\omega_2) \, E_\beta(\omega_1)
E_\gamma(\omega_2) \\ \nonumber
&&\!\!\!\!\!\!\!\!\!\!\!\!\!\!\!\!\!\!\!\!\!\!\!\! + \frac{1}{2}
\, \gamma_{\alpha \beta
\gamma\delta}(\omega_3=\omega_1+\omega_2+0) \, E_\beta(\omega_1)
E_\gamma(\omega_2) \, E_\delta(0) \; ,
\end{eqnarray}
where $\beta_{\alpha \beta \gamma}$ and $\gamma_{\alpha \beta
\gamma \delta}$ are the first and second hyperpolarizability
respectively. The macroscopic response is given by an appropriate
spatial average. Here we consider an isotropic medium in the
presence of a static electric field $E_\delta(0)$. A Boltzmann
average yields the induced macroscopic polarization at the
sum-frequency \cite{pfcpl98,pfacs01a}:
\begin{eqnarray}
\label{polsfg} P_\alpha(\omega_3)&=& \epsilon_0 \,
E_\beta(\omega_1) E_\gamma(\omega_2) \; \Bigg[
\overbrace{\epsilon_{\alpha \beta \gamma} \;
\chi^{(2)}}^{\mathrm{chiral}}  \\  && \!\!\!\!\!\!\!\!\!\!\!\!\!\!
+ \underbrace{3\, \Big( \chi^{(3)}_1
\delta_{\alpha\beta}\delta_{\gamma\delta} + \chi^{(3)}_2
\delta_{\alpha\gamma}\delta_{\beta\delta} + \chi^{(3)}_3
\delta_{\alpha\delta}\delta_{\beta\gamma} \Big)
E_\delta(0)}_{\mathrm{achiral}} \Bigg] \; , \nonumber
\end{eqnarray}
where $\epsilon_{\alpha\beta\gamma}$ is the Levi-Civita tensor and
$\delta_{\alpha\beta}$ the Kronecker delta. $\chi^{(2)}$ is, as we
will show, the completely antisymmetric chirally sensitive
isotropic component of the second-order susceptibility, and
$\chi^{(3)}_i$ with $i=$1, 2 or 3 are the achiral isotropic
components of the third-order susceptibility. The intensity at the
sum-frequency is proportional to $\left| P_\alpha(\omega_3)
\right|^2$ and so contains -- besides a contribution independent
of the static field, and one that is quadratic in the static field
-- cross terms that are linear in the static field and linear in
$\chi^{(2)}$:
\begin{equation}
{\small \textrm{SFG(E)}} \propto \mathrm{Re}\!\left[\chi^{(2)} \;
\left(\chi^{(3)}_i\right)^* \, \right] \; E(0) \; I(\omega_1) \,
I(\omega_2) \; ,
\end{equation}
where $I$ are the respective incident intensities, Re is the real
part and the star indicates a complex conjugate. We have denoted
the contribution to the intensity at the sum-frequency that is
linear in the static field by SFG(E).

$\chi^{(2)}$ is proportional to the isotropic component of the
first hyperpolarizability and is given by
\begin{eqnarray}
\label{chi2} \chi^{(2)}= \frac{N}{12\, \epsilon_0}
\left(\beta_{xyz} - \beta_{xzy} + \beta_{yzx} - \beta_{yxz} +
\beta_{zxy} - \beta_{zyx} \right) \, , \nonumber
\end{eqnarray}
where $N$ is the excess number density.
 It is seen that $\chi^{(2)}$ vanishes for any molecule
that possesses a mirror plane, a center of inversion, or a
rotation-reflection axis; thus $\chi^{(2)}$ is only non-zero for
chiral systems and changes sign as the handedness of the optically
active liquid changes. The contribution to the intensity linear in
$E(0)$ may therefore reveal the sign of the pseudoscalar
$\chi^{(2)}$. It can be used to determine the absolute
configuration of the chiral molecules in the optically active
liquid.

$\chi^{(3)}_1, \chi^{(3)}_2$ and $\chi^{(3)}_3$ are given in terms
of scalar combinations of the second hyperpolarizability tensor
components by \cite{pfcpl98}:
\begin{eqnarray}
\chi_1^{(3)} &=& \frac{N}{180 \: \epsilon_0} \left[+ 4
\gamma_{\alpha \alpha \beta \beta} -
\gamma_{\alpha \beta \alpha \beta} - \gamma_{\alpha  \beta \beta \alpha} \right] \nonumber \\
\chi_2^{(3)} &=& \frac{N}{180 \: \epsilon_0} \left[ -
\gamma_{\alpha \alpha \beta \beta} + 4
\gamma_{\alpha \beta \alpha \beta} - \gamma_{\alpha  \beta \beta \alpha} \right] \nonumber \\
\chi_3^{(3)} &=& \frac{N}{180 \: \epsilon_0} \left[ -
\gamma_{\alpha \alpha \beta \beta} - \gamma_{\alpha \beta \alpha
\beta} + 4 \gamma_{\alpha  \beta \beta \alpha} \right] \, .
\end{eqnarray}
Should the duration of the static applied field be longer than the
rotational time of the molecules in solution, then there are
additional temperature-dependent terms of the form
$\beta_{\alpha\alpha\beta}(\omega_3=\omega_1+\omega_2) \,
\mu^{(0)}_\beta /(kT)$  for molecules with a permanent dipole
moment $\mu^{(0)}_\beta$ \cite{pfcpl98}. At ambient conditions,
the temperature dependent contributions may dominate for dipolar
molecules and be of either sign. Both $\gamma_{\alpha \alpha \beta
\beta}$ and $\beta_{\alpha\alpha\beta}\,\mu^{(0)}_\beta$ are
unchanged under mirror symmetry operation and so it follows that
the $\chi^{(3)}_i$ exist for achiral molecules (e.g. an achiral
solvent) and are necessarily the same for both enantiomers of a
chiral molecule.

\begin{figure}[ht]
\centerline{\includegraphics[width=8cm]{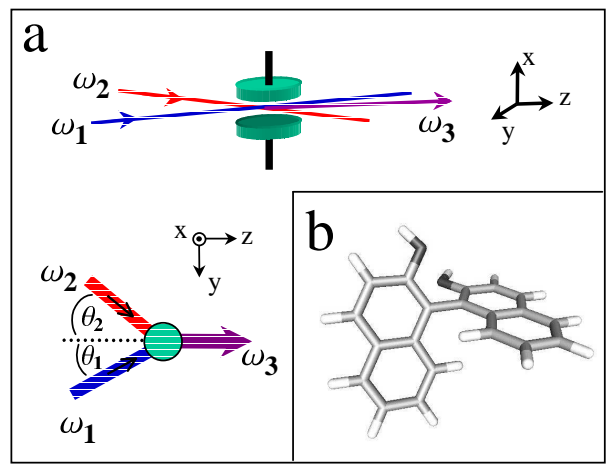}}
\caption{\label{fig1} (color online) a) Beam geometry and choice
of axes. The optical beams lie in the $y,z$ plane, and the static
field is S-polarized along $x$. The incident beam at $\omega_1$
makes an angle of $\theta_1$ with the sum-frequency beam at
$\omega_3$; and similarly for $\omega_2$. (The cuvette with the
optically active liquid is not shown.) b) Structure of
R-$(+)$-$1,1'$-bi-2-naphthol (C$_{20}$H$_{14}$O$_2$).}
\end{figure}

We observe the chiral electro-optic effect in an experimental
arrangement schematically depicted in Figure \ref{fig1}a. From the
antisymmetry of the Levi-Civita tensor in Eqn. (\ref{polsfg}) it
follows that the optical fields need to have components that are
mutually orthogonal such that they span the $x$, $y$, and $z$ axes
of a Cartesian frame. Hence, sum-frequency generation from
isotropic media requires that one optical field be S-polarized and
the remaining two be P-polarized. Further, the static electric
field has to be S-polarized for it to give rise to an
electric-field-induced contribution to the sum-frequency signal.

We use the 775-nm fundamental wavelength of a Ti:sapphire
regenerative amplifier (Clark-MXR CPA-2010) along with its second
harmonic at 338 nm to generate sum-frequency signals at 258 nm.
The pulse duration is $\sim$150 fs and the repetition rate is 1
 kHz. A custom-built high voltage power supply provides 3.5 to 10
kV peak to peak modulated at 500 Hz. The modulation of the applied
electric field permits phase sensitive detection.

Unlike in collinear electric-field-induced second-harmonic
generation (EFISH) \cite{levine}, the non-collinear SFG beam
geometry \cite{pfprl00} allows the use of fully immersed
electrodes. These are suspended in a standard quartz optical
cuvette and spaced $\sim$2 mm apart.

\begin{figure}[ht]
\centerline{\includegraphics[width=8cm]{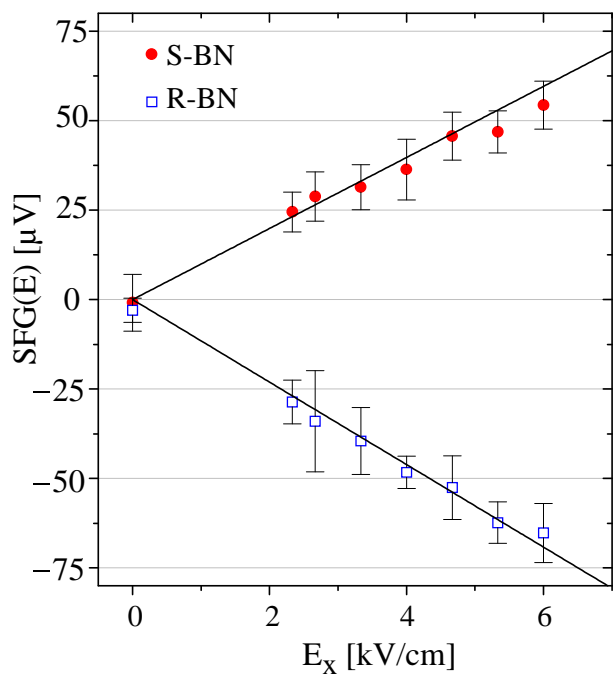}}
\caption{\label{fig2} (color online) SFG(E) signals from $\sim$0.5
M solutions of S-$(-)$- and R-$(+)$-$1,1'$-bi-2-naphthol
respectively measured at different strengths (estimated) of the
applied electric field. The solid lines are linear functions
fitted to the data.}
\end{figure}

We observe the chiral electro-optic effect in optically active
solutions prepared from R-$(+)$- or S-$(-)$-$1,1'$-bi-2-naphthol
(R-BN and S-BN) (see Figure \ref{fig1}b) which are dissolved in
the achiral solvent tetrahydrofuran (THF). In Figure \ref{fig2} we
show that the signals are of opposite sign for the R- and
S-enantiomers of BN and that they vary linearly with the strength
of the applied low-frequency field. The small difference in
absolute strength of the respective signals is attributed to small
differences in the electrode-spacings and/or the concentrations.
The SFG(E) signals change sign as the direction of the static
field is reversed.

\begin{figure}[ht]
\centerline{\includegraphics[width=8cm]{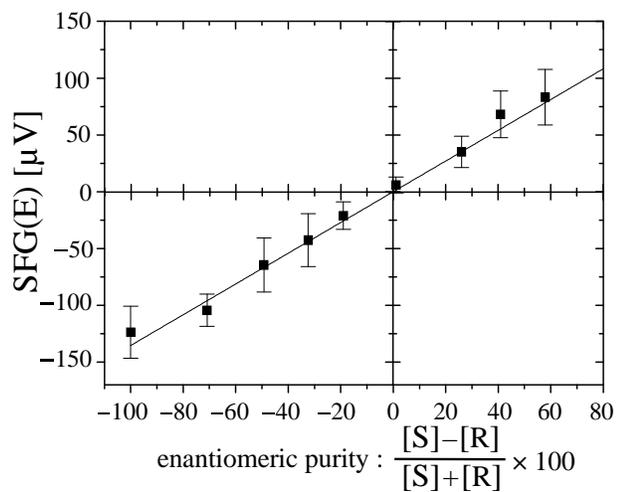}}
\caption{\label{fig3} SFG(E) signals measured as a function of the
fractional concentration difference of R-($+$)- and
S-($-$)-$1,1'$-bi-2-naphthol in THF. The solid line is a linear
function fitted to the data.}
\end{figure}

For fixed beam polarizations, direction and strength of the static
field, we also observe a linear dependence of the chiral
electro-optic effect on the enantiomeric concentration difference
of the optically active solution, as is seen in Figure \ref{fig3}.
Starting with a 0.56 M solution of R-$(+)$-$1,1'$-bi-2-naphthol in
THF, the handedness of the 0.56 M solution is gradually changed by
the addition of a solution of S-$(-)$-$1,1'$-bi-2-naphthol in THF.
The SFG(E) signal correspondingly changes sign. It is zero for the
racemic mixture.

To describe the polarization dependence of the sum-frequency
signals, we now consider the axis system and beam geometry
depicted in Figure \ref{fig1}a. For simplicity we ignore the small
rotation of the plane of polarization in the chiral liquid due to
optical activity ($<0.5^\circ$), and we note that a static
electric field can not contribute linearly to optical rotation in
a fluid.

\begin{figure}[ht]
\centerline{\includegraphics[width=8cm]{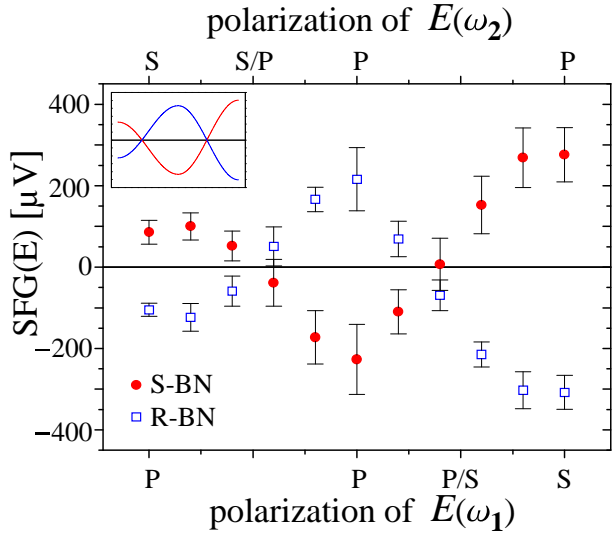}}
\caption{\label{fig4} (color online) Sum-frequency generation
signals from optically active $\sim$0.5 M solutions of S-($-$)-
and R-($+$)-BN in THF observed for different beam polarizations.
The polarization state of the $\omega_1$ beam is shown on the
lower axis and that of the $\omega_2$ beam on the upper axis.
First, $E(\omega_1)$ is kept P-polarized while the polarization of
$E(\omega_2)$ is changed from S to P in 9$^\circ$ increments.
Subsequently, $E(\omega_2)$ remains P-polarized while the
polarization of the $\omega_1$ beam is changed from P to S. When
both the incident optical fields at $\omega_1$ and $\omega_2$ are
P-polarized, then the $\omega_3$ beam is S-polarized. Mixed input
polarizations (S and P) give rise to a P-polarized signal. The
static electric field $E_x$ is  $\sim$ 2 kV/cm. The expected trend
of the signals is shown by a model calculation depicted in the
inset.}
\end{figure}

\begin{figure}[ht]
\centerline{\includegraphics[width=8cm]{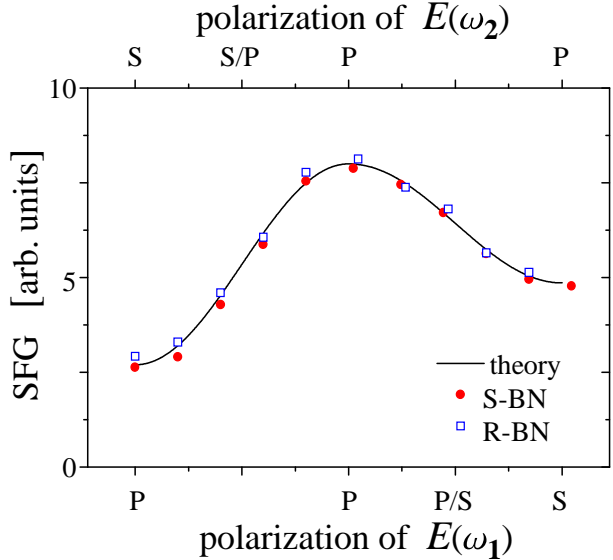}}
\caption{\label{fig5} (color online) The SFG intensity in the
absence of a static electric field is shown as a function of the
input beam polarizations (see caption Fig.~\ref{fig4} for
details). The solid line is a fit from theory. The contribution
quadratic in the static field is below the sensitivity of the
experiment.}
\end{figure}

From Eqn. (\ref{polsfg}) and the requirement for a transverse wave
at the sum-frequency, it follows that a PPS combination of
polarizations (listed throughout this Letter in the order of the
optical fields at: $\omega_3$, $\omega_1$, $\omega_2$) gives rise
to an SFG(E) signal of the form
\begin{eqnarray}
{\small \textrm{SFG(E)}}_\mathrm{PPS} = (+) \, \eta \,
\mathrm{Re}\!\left[ \chi^{(2)} \left(\chi^{(3)}_1\right)^*
\,\right] \sin(2\theta_1) \, E_x(0) \,, \nonumber
\end{eqnarray}
where $\theta_1$ is the angle the incident beam at $\omega_1$
makes with respect to the sum-frequency beam, and where we have
subsumed the incident intensities and any numerical factors common
to all SFG(E) intensities in the positive multiplier $\eta$.
Similarly, PSP polarizations give rise to
\begin{eqnarray}
{\small \textrm{SFG(E)}}_\mathrm{PSP} = (+) \, \eta \, \mathrm{Re}
\!\left[ \chi^{(2)} \left(\chi^{(3)}_2\right)^*\,\right]
\sin(2\theta_2) \, E_x(0) \,, \nonumber
\end{eqnarray}
and SPP polarizations probe
\begin{eqnarray}
{\small \textrm{SFG(E)}}_\mathrm{SPP} = (-) \,\eta \,
\mathrm{Re}\!\left[\chi^{(2)}\left(\chi^{(3)}_3\right)^* \,\right]
\times && \nonumber  \\ && \!\!\!\!\!\!\!\!\!\!\!\!\!\!\!\!\!\!
\sin(2(\theta_1+\theta_2)) \, E_x(0) \,. \nonumber
\end{eqnarray}
Since $\chi^{(3)}_i$ are the same for the R- and S-enantiomers
while the $\chi^{(2)}$ change sign, solutions of R- and S-BN with
equal concentration will have equal but opposite SFG(E)
intensities in the same field $E_x(0)$, and this can be seen in
Figure \ref{fig4}. Further, the change of sign in going from PPS
to SPP polarizations and similarly from SPP to PSP suggests that
$\chi^{(3)}_1, \chi^{(3)}_2$ and $\chi^{(3)}_3$ have the same sign
for BN ($2(\theta_1+\theta_2)<\pi$). The three polarization states
 PPS, PSP and SPP permit the observation of the three isotropic
third-order susceptibilities, $\chi^{(3)}_i$. For a static field
of $\sim$2.5 kV/cm the SFG(E) signals are, depending on beam
polarizations, $\sim$5 to 20\% of the SFG signals in the absence
of a dc electric field shown in Figure \ref{fig5}. Nonlinear
optical susceptibilities of different order may thus be measured
under identical optical field conditions without the need for an
external reference, as the $\chi^{(2)}$ will act as an internal
standard. The relative strength of $\chi^{(3)}_1, \chi^{(3)}_2$
and $\chi^{(3)}_3$ may in turn yield information about specific
hyperpolarizability tensor components entering the third-order
susceptibility. Such an analysis is, however, chromophore specific
and will not be discussed further here. Finally, Figure \ref{fig5}
confirms that SFG signals independent of $E$ do not distinguish
between enantiomers.

For a dc electric field along $+x$ (see Fig \ref{fig1}a) we
measure a positive $\mathrm{SFG(E)_{PSP}}$ intensity for S-BN. In
the case of the resonant signals from R- and S-BN, where the
complex nature of the response tensors needs to be considered, and
in the absence of any measurable $\chi^{(3)}_i$ contribution from
the achiral solvent, the determination of the absolute
configuration of the chiral molecules will require quantum
chemical computations for both $\chi^{(2)}$ and $\chi^{(3)}_i$.
However, the contribution from an achiral solvent should make it
possible to determine the absolute sign of both the second- and
third-order susceptibilities for both optical isomers of a chiral
molecule \cite{pfcpl98}.

In summary, we have observed the linear effect of a static
electric field on the sum-frequency generation intensity from
optically active liquids. The reported chiral electro-optic effect
arises in the electric-dipole approximation and changes sign with
the enantiomer and upon reversing the direction of the static
electric field. The effect may be used to determine the absolute
sign of the second-order and third-order nonlinear optical
susceptibilities.

Should the achiral solvent give rise to an appreciable
electric-field-induced sum-frequency signal, then the effect could
be used to amplify weak sum-frequency signals from chiral
solutions.

The authors are grateful to Professor A.C. Albrecht for many
helpful discussions, and thank Yi-Fan Chen for assistance with the
experiment. Clark-MXR is acknowledged for the use of the
regenerative amplifier. This work was supported by the National
Science Foundation (CHE-0095056) and the National Institutes of
Health (EB002019).

\end{document}